# H²CM-based holonic modeling of a gas pipeline

C. Indriago*, L. Ghomri**, O. Cardin***

*Universidad Politécnica Antonio José de Sucre.
(CENIPRO Centro de Investigación de Procesos)
Barquisimeto-Venezuela.

(e-mail: cindriago@unexpo.edu.ve)

**Manufacturing Engineering Laboratory of Tlemcen
Department of Electrical Engineering and Electronics
Abou-Bekr Belkaïd University
Tlemcen – Algeria

(e-mail: ghomri@mail.univ-tlemcen.dz).

***LUNAM Université, IUT de Nantes – Université de Nantes, LS2N UMR CNRS 6004
(Laboratoire des Sciences du Numérique de Nantes),
2 avenue du Prof. Jean Rouxel – 44475 Carquefou

(e-mail: olivier.cardin@ls2n.fr).

**Abstract.** A gas pipeline is a relatively simple physical system, but the optimality of the control is difficult to achieve. When switching from one kind of gas to another, a volume of useless mixture is generated. Therefore, the control needs to both respond to the demand and minimize the volume of lost gas. In case of stable and perfectly known demand, scheduling techniques can be used, but in other cases, calculation times are incompatible with an industrial application. This article introduces the application of H²CM (Holonic Hybrid Control Model) generic architecture on this specific case. The study case is extensively presented. Then, the defined holonic architecture (H²CM compatible) is detailed, and the role and functions of each holon are presented. Finally, a tentative general control algorithm is suggested, which gives an insight on the actual algorithms that will be developed in perspective of this work.

# 1   Introduction

Hybrid dynamic systems (HDS) are dynamic systems integrating explicitly and simultaneously continuous systems and discrete event systems. They require for their description the use of continuous time models, discrete event models and the interface between them [1]. The hybrid character of the system either owes to the system itself or to the control applied to this system. Typical examples of such systems are communication protocols, manufacturing systems, transportation systems, power electronics, etc.

Modeling, analysis and control of HDS are crucial concerns. Two of the most important formalisms for the modeling and analysis of HDS are hybrid automata [2] and hybrid Petri nets [3]. Hybrid automata can consider any continuous dynamics in a location, and the commutation from one location to the other one is synchronized by a discrete event. It is then possible to model any type of system. The main modeling drawback of the hybrid automata is the explosion of the number of locations in case of real life systems. To overcome this problem, hybrid Petri nets consider a state as a marking with a continuous and a discrete part. They provide very compact and readable models very useful for engineers. However, in order to perform a formal analysis, it is necessary to come back to the hybrid automata which are known for their analysis power. This analysis requires the construction of the reachable state space. This operation is all the more complex because the discrete part is strongly nonlinear and the time is often non-deterministic. The calculation algorithms of the reachable state space only terminate under very restrictive constraints, for example for timed models or some linear hybrid automata where the continuous dynamics are constant. The reachable state space is described with a set of inequalities over the state variables, thus allowing both the performance analysis of the system and the synthesis of the control.

Most realistic formal approaches consider either continuous approaches with few commutations or discrete approaches with a very poor continuous dynamics (clocks). Since these systems are strongly nonlinear, any change in one or several parameters often forces us to completely redo the study of the problem. When the HDS become more complex, the analysis tools also become more complex, turning them into loosely flexible systems with high calculation times, which do not react fast enough to unexpected events. In order to provide flexibility to the HDS, the researchers have studied the possibility of implementing flexible control architectures on complex dynamic systems [4][5].

Holonic Hybrid Control Model (H²CM) [5] is a holonic architecture developed with the aim of giving flexibility to HDS control and it is based on the holonic architecture of discrete systems called PROSA [6]. It is composed of three basic holons:

- the product holon, which has all the information of the product;
- the resource holon, which is an abstraction of the resource;

- the order holon, which takes the information of both holons and generates the scheduling of the services to implement.

An example of a complex HDS is the pipeline system [7], which is a very important transport system that guarantees a regular supply of products and a rapid adaptation to market demand, thus significantly reducing costs and delays of product transportation. Pipeline systems are continuous operation systems that work with several products, resulting in a contaminated mixing zone on the contact of two products that are transported sequentially. Therefore, a greater number of batches transported will produce a larger number of contaminated product batches. With the objective of minimizing contaminated product batches, optimization methods are used that generate the batch sequence scheduling to be transported, with the restriction that these methods have high calculation times and are not flexible to changes in transport demand.

The objective of this article is to propose the implementation of the H²CM architecture to a pipeline transport system in order to provide operational flexibility in the phase of changes in the demand for products, keeping the optimization criteria of the generation of contaminated product. This preliminary study extends the performance evaluation of H²CM that was made on a water tanks system [8] with the notion of switching costs (contaminated product volume) and the dynamics of the pipeline (delay between the switch and the final tank filling).

The study case is extensively presented in the next section. Then, the defined holonic architecture (H²CM compatible) is detailed, and the role and function of each holon are presented. Finally, a tentative general control algorithm is suggested, which gives an insight on the actual algorithms that will be developed in perspective of this work.

## 2  Case study: pipeline presentation

### 2.1  Description of the ASR multi-product pipeline

The transportation of fuels by pipeline is increasingly spread throughout the world. This is explained by an increase in the quantities of transported products. This situation requires companies to further develop their logistics. It is in this objective that the Algerian oil companies have an investment program, aimed at securing the country's petroleum products, through an intelligent network of pipelines, responding to the real need of the different zones of the country.

The transportation by pipeline contributes to the reduction of both costs, delivery times, road traffic and also ensures mass transport respecting the environment with the most security. If pipelines did not exist, it would be inevitable to have thousands of trucks and railcars that circulate on roads, highways and railways to carry out the same transport.

The current technology is oriented towards multi-product pipelines. The latter has the disadvantage of creating a mixing or a contaminated product zone between two products in contact (Fig. 1), which circulate sequentially in the pipeline. The contaminated product is generated at each contact of two different products of fuels. So a sequence of several batches promotes proportionately several batches of contaminated product, requiring a large space for their storage.

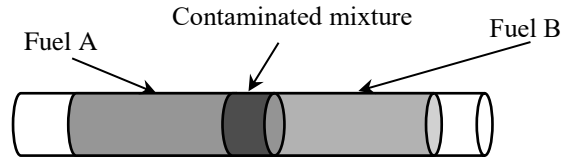

**Fig. 1.** Typical sequence with two products and one contaminated area

In this paper, we focus our study on a typical pipeline of the National Petroleum Algerian Company (SONATRACH). It is the multi-product ASR pipeline (Abbreviation for the three cities: Arzew, Sidi Bel Abbès and Remchi in the west of Algeria, through which the pipeline passes). The pipeline transports fuel from the Arzew refinery to the storage and distribution sites of Sidi Bel Abbes and Remchi. They feed the regions West and South-West of Algeria in fuels (diesel fuel and gasolines). The demand of the region, leads to the introduction into the ASR pipeline of important sequences of several batches to meet the needs of the region. This results in numerous interfaces, zones of birth of the mixtures and high levels of the contaminated product stock at the depot of Remchi.

In view of the contamination constraints and the high demand of fuel, it will be interesting to study the optimization of the multi-product fuel transport. The objective will therefore be the minimization of the contaminated product stocks on one hand, and the satisfaction of the of the demands of the two distribution depots of pure products on the other hand,

## 2.2 *Physical data of the pipeline:*

The multi-product ASR pipeline is located in the west of Algeria. Its profile extends over a length of about 168 *km*, from the refinery of Arzew passing by Sidi Bel Abbes depot and arriving at the final depot of Remchi. The pipeline receives the liquid fuels from the Arzew refinery and supplies the storage and distribution depots in Sidi Bel Abbes and Remchi (Fig. 2).

Table 1 below shows the storage capacities of each depot in the different fuels. In table 2 we represent the daily demand at the two depots level.

## 3 Holonic Modelling

### 3.1 H²CM overview

H²CM generic architecture is based on the three basic holons of PROSA, introduced in Fig. 4 (a). Two main features can be highlighted:

1. Each resource is granted with an order and a product along its life. The order holon is in charge of the monitoring of the resource whereas the product holon is in charge of the recipe to be applied on the actual product. The content and objectives of the order and product holons are constantly evolving, but the structure remains constantly the same;
2. A recursivity link is present on the resource holon (Fig. 4 (b)). Indeed, each compound resource can be fractally decomposed into one or several holarchies, comprising one or several resources and their associated order and product holons. The aggregation relations created here can be changed along the working of the system; holarchies can be created and destroyed online.

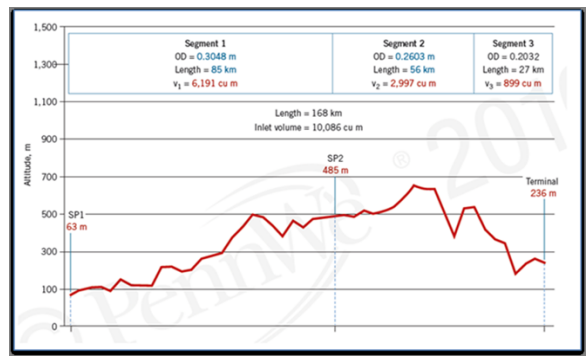

**Fig. 2.** Longitudinal profile of the multi-product pipeline ASR.

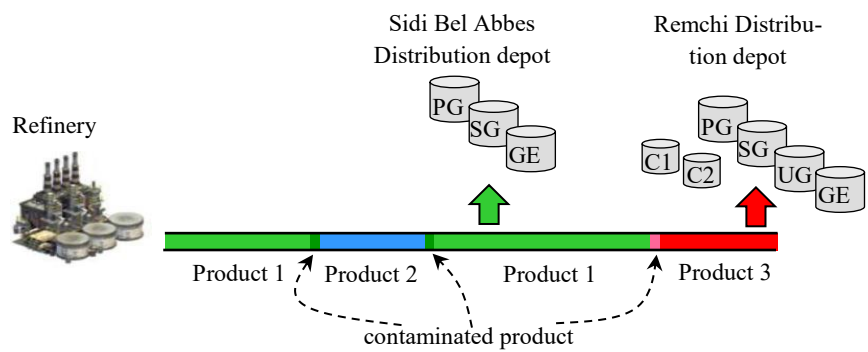

**Fig. 3.** Structure of the ASR structure.

**Table 1.** Capacities, initial volumes and safety stocks of products and contaminated products in the two depots.

| Fuels | Sidi Bel Abbes depot | | | Remchi depot | | |
|---|---|---|---|---|---|---|
| | Capacity [m³] | Initial stock [m³] | Security stock [m³] | Capacity [m³] | Initial stock [m³] | Security stock [m³] |
| **Diesel** | 6000 | 814 | 1200 | 22000 | 5572.4 | 4400 |
| **Pure gasoline** | 1700 | 809 | 340 | 9500 | 3394.6 | 1900 |
| **Super gasoline** | 450 | 196 | 90 | 1000 | 996 | 200 |
| **Unleaded gasoline** | | | | 5000 | 3284.7 | 1000 |
| **Contaminated product type 1** | | | | 500 | 396.5 | |
| **Contaminated product type 2** | | | | 500 | 405 | |

**Table 2.** Daily demand in the two depots.

| Product | Daily demand | |
|---|---|---|
| | Sidi Bel Abbes depot | Rechi depot |
| **Diesel** | 1200 | 3000 |
| **Pure gasoline** | 80 | 150 |
| **Super gasoline** | 400 | 800 |
| **unleaded gasoline** | | 150 |

The order holon is globally responsible for the scheduling of the system. It is very representative of the dual working of every holons: one part is dedicated to the negotiation mechanisms in look-ahead mode in order to determine the future scheduling and holarchies of the system, while the other part of the holon is dedicated to the system in real time mode through the application of the scheduling resulting from the previous negotiations.

The product holon is dedicated to store and communicate the products recipes which are deduced from a Master Recipe. In H2CM, this Master Recipe can be defined as a generic recipe, i.e. a sequential list of operations to be applied to the product to obtain a final product from raw materials (BOM – Bill of Materials), from which the actual recipe can be derived according to the conditions of the system. In HDS context, a service-oriented specification, as proposed in [9, 10], is well suited for the product specification. The distinction made in this article, with respect to the definition used in [9, 10], is that the parameters and variables of the service can be continuous or discrete.

By nature, HDS are large systems, constituted of many components. Therefore, a lot of resource holons are necessary to control the system. As Fig. 4 (b) defined a recursivity of the resources, so the smallest resource holon to be defined is called atomic resources and can be expressed as "the maximal aggregation of elements whose system of differential equations can be inversed in a short delay relatively to the dynamics of the system". Considering compound resources, the negotiation mechanism is meant to determine the best solution recursively, interrogating the aggregated resources until atomic ones. Resources holons are part of the negotiation mechanisms and their function is to evaluate and transmit to the order holon the best possible variables values to obtain the desired function and services. Resource holons also have the responsibility for the devices online control, i.e. the role of controllers of the system.

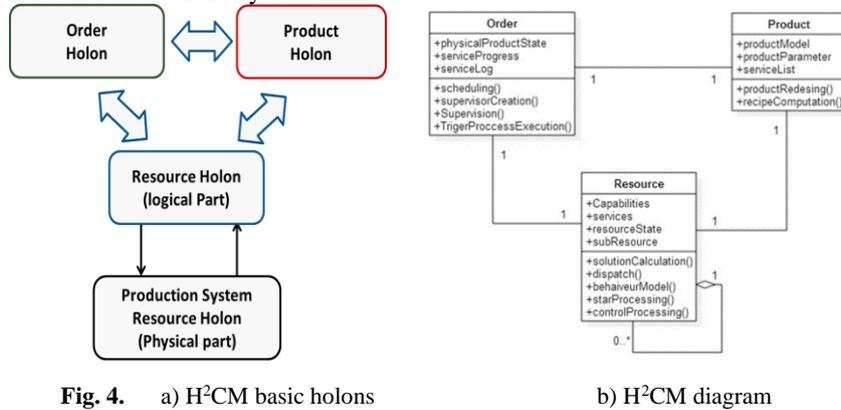

**Fig. 4.**   a) H$^2$CM basic holons             b) H$^2$CM diagram

Another resource holon's specificity is its structure. Classically, it is composed by a physical part and a logical part, see Fig. 4 (a). The physical part is represented by the shop floor. The logical part of the resource holon is an abstraction of the physical part and contains the conversion models from continuous states to discrete states and vice-versa. The models used are hybrid models that change their state using threshold levels of continuous variables. The abstraction of physical part can be developed by any industrial computer with communication capability.

### 3.2   Description of holons and services in the case of a pipeline

The work developed in this paper focuses on the multi-product ASR pipeline linking the Arzew refinery to the storage and distribution centers of Sidi Bel Abbes and Remchi. The latter deserves the centers in fuels namely: Diesel, Super Gasoline, Unleaded Gasoline and Normal Gasoline. From the holonic point of view, the ASR pipeline will be divided into three composite holons (see Fig. 5) which will offer three different types of services.

The first composite holon to be defined is the Arzew refinery, composed by the finished product tanks system and the pipe and pump system. The service offered by the holon of the Arzew refinery is the supply of fuel products. Looking at Fig. 6, the refinery resource holon has an associated product holon and an order holon. The product holon has contaminated product information, such as the volume generated in each product mix and the variation of product density during mixing. The order holon will have the task of scheduling and executing the product supply service. The order holon will perform a scheduling using off-line optimization methods, so it will need the atomic resources holon information and the store holons information, obtained by holons negotiation. A second task of the order holon is based on the supervision of real-time scheduling, if any disturbance occurs in the execution of the same, the holon order must take corrective measures with the new information and perform a re-scheduling, this procedure will be discussed in the next section.

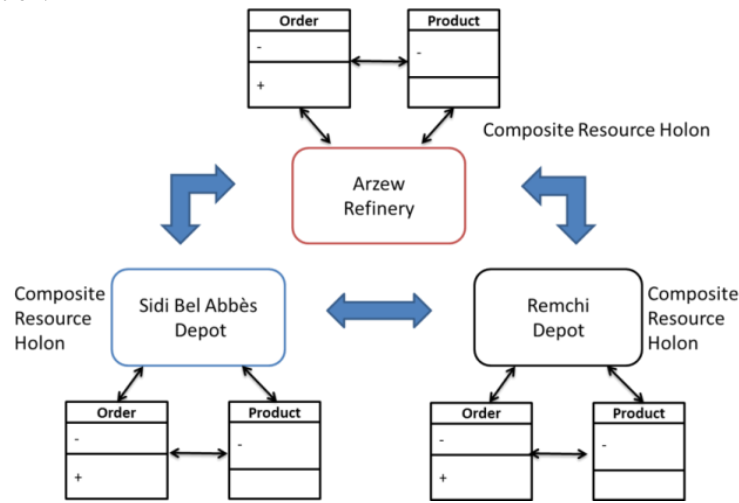

**Fig. 5.** Distribution level holonic architecture

On the other hand the refinery resource holon is composed of tank holons, which for the case under study is considered an inexhaustible supply. Therefore, its function is to switch the supply valve depending on the product to be supplied. If there was no inexhaustible supply, this holon should supply the level status of each product to be considered during the scheduling. The other holon available in the refinery holon is the pipeline holon, its task is based on determining the availability of the pipeline as well as providing the products density measure service and the control pumps service during product supply.

The other two holons represented in Fig. 5 are the products depots holons of Sidi Bel Abbès and Remchi. Both composite holons are made up of diesel tanks, Super gasoline tank, and normal gasoline tank, this latter depot additionally has unleaded gasoline tank and contaminated product tank. Both depots offer the ser-vice of storage of finished product (normal gasoline, super gasoline and diesel). In addition, Remchi depot offers the storage service of unleaded gasoline and contaminated product, see Fig. 7.

Each holon depot also has associated a product holon and an order holon. The information possessed by the product holon is the characteristic of each products to be received, among them we must highlight the products density since this information is used to monitor the products in the pipeline. The order holon contains information on the capacity of each tank resource holon. The method used to obtain this information from the compound holons is detailed in [5], [11]. The service provided by the order holon is based on online monitoring filling of the product tanks. If there is a change in the demand, it establishes a new negotiation with the refinery holon to obtain a corrective action.

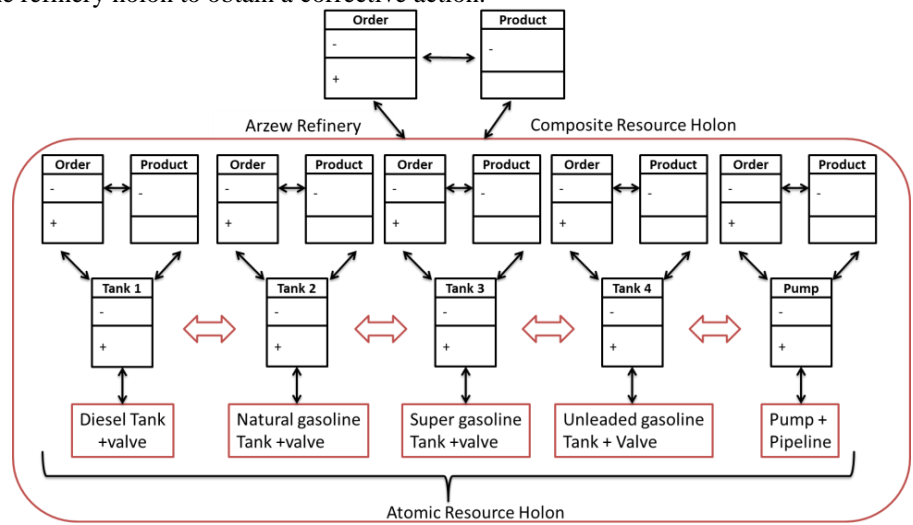

**Fig. 6.** Refinery Composite Holon

This description of the adaptation of H²CM reference architecture to this new case study outlines the adaptability of the architecture, that was intended generic enough to cope with multiple type of HDS.

## 4  Algorithms

The main goal of the product supply scheduling algorithm is to minimize the amount of contaminated product while supplying the tanks to ensure the demand. For this, it is necessary to execute an optimization method that guarantees the desired objective. Usual optimization methods have high calculation times which make them difficult to implement online. Thus, the system can perform a first schedule using optimal methods since it has sufficient time to perform the calcula-tion, but this schedule needs to be updated in real time using other techniques. This reschedule is obtained by negotiating between the refinery holon and the depot holons.

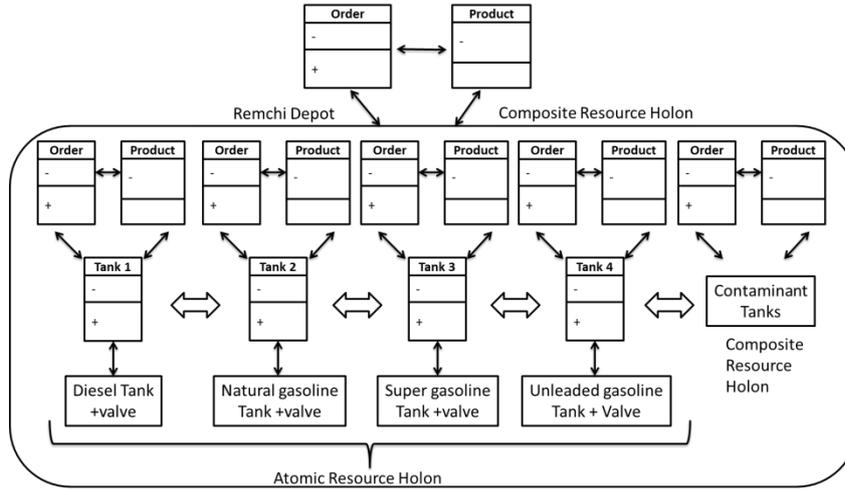

**Fig. 7.** Depot Composite Holon

Once the schedule is calculated (by any method), the scheduling supervision process begins. Once an unforeseen event is present in the product supply process, the order holon needs to make a decision based on the information obtained through the negotiation of the holons the resolve this new disturbance. It is obviously difficult to exhaustively list all the disturbances that may occur in the system, but the following ones can be considered as the most frequent and impacting:

1. Variation in demand for products depots;
2. Variation of the tanks capacity of contaminated product;
3. Decreased pump performance of the pipeline system.

These disturbances, when occurring on a system controlled by a schedule based on optimal methods without adjustments in line, can cause changes in the system that result in a shutdown of the functions in the transport system. For this, an online scheduling algorithm is proposed, that aims to find a solution to the presented perturbation, see Fig. 8.

In Fig. 8 is observed that once the disturbance is presented, the incidence on the current planning is calculated as a function of time. If the present disturbance is not so aggressive that immediate changes are needed, then re-scheduling will be performed by optimal methods, otherwise re-scheduling will be done by any other computational method that is fast enough to find a solution. An online schedule algorithm for tanks filling minimizing the number of switches of the unique server was proposed in [8], and might be adapted to this specific case study. The main idea is to try and maximize the duration the server remains in the same position, while avoiding the situation where the rest of tank are empty at the same time.

Therefore, the algorithm tries to anticipate the reservation of the server if the time slot where some tank would be empty is already scheduled to another tank. A timelapse is also defined. This timelapse represents the length of the reservation a tank tries to schedule. If no solution can be found by the system with the predetermined timelapse, then the algorithm loops with a shorter timelapse. The calculation time of the algorithm is very short, which makes us believe that this kind of scheduling, although it does not guarantee the best minimization of the amount of generated contaminated product, is a good candidate for a pertinent control in the situations where the input data are frequently disturbed.

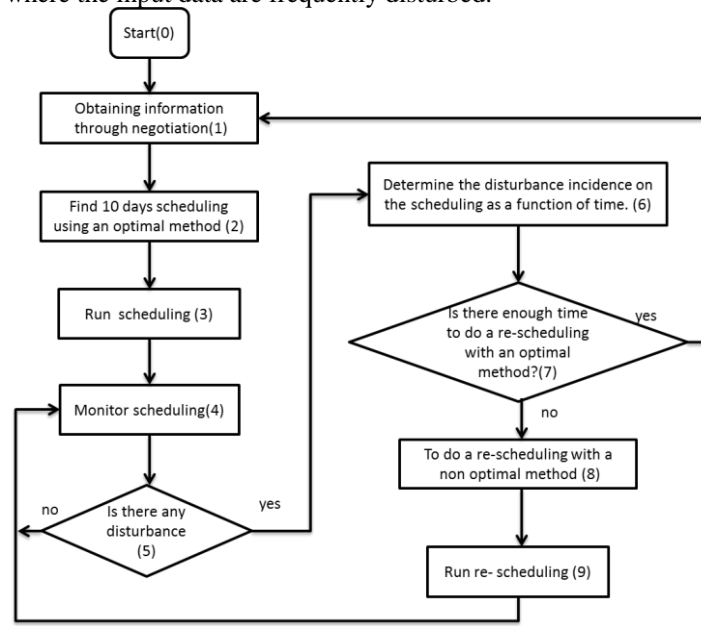

**Fig. 8.** Scheduling and re-scheduling Algorithm

The algorithms that need to be developed in this case study greatly differ from the ones previously published [8]. Indeed, the former was related to a single server of water and no setup times, while this is related to a single server generating setup wastes denoted contaminated product in this description.

## 5    Conclusion

Pipeline transport systems are complex HDS that can be scheduled by optimal methods, but they have little flexibility facing disturbances, which makes them difficult to control using online methods. To this end, the $H^2CM$ implementation on HDS was previously proposed [5] in order to give flexibility to those systems.

To demonstrate the application of H$^2$CM on HDS, a case study of the multi-product ASR pipeline was used in this article. This preliminary work outlined the adaptation of H²CM to this specific case study and showed it was suitable for modelling. Some elements were given about the definition of the future algorithms that will control the whole architecture, with the objective of finding solutions through optimal and non-optimal methods, and also allowing online monitoring and re-scheduling of the system in presence of any deviation.

The perspectives of this preliminary work deal with the actual coding of the architecture and the performance evaluation compared to the scheduling techniques with a perfectly known demand in order to evaluate the optimality of the holonic control. Then, a study will be performed in order to evaluate the robustness of the control in case of demand variation. The H²CM based control is meant to absorb those uncertainties in real time, which needs to be verified.